\documentclass[a4paper,11pt]{article}
\usepackage{pos}
\pdfoutput=1   
\usepackage{mathrsfs,latexsym}
\usepackage{graphicx}
\usepackage{xcolor}
\usepackage{booktabs}
\usepackage{multirow}
\usepackage{multicol}
\usepackage{appendix}
\usepackage{fontawesome} 
\usepackage{bm}
\usepackage{times}
\usepackage{float}
\usepackage{wrapfig}
\usepackage[utf8]{inputenc} 
\usepackage[T1]{fontenc} 
\usepackage{hyperref}

\newcommand{\Eq}[1]{Eq.~(\ref{#1})}

\newcommand{\fig}[1]{figure~\ref{#1}}

\newcommand{\tab}[1]{table~\ref{#1}}

\newcommand{\tr}{\mbox{Tr}}

\newcommand{\ben}{\begin{enumerate}}
\newcommand{\een}{\end{enumerate}}
\newcommand{\bit}{\begin{itemize}}
\newcommand{\eit}{\end{itemize}}
\newcommand{\beq}{\begin{equation}}
\newcommand{\eeq}{\end{equation}}
\newcommand{\bsa}{\begin{subequations}\begin{eqnarray}}
\newcommand{\esa}{\end{eqnarray}\end{subequations}}
\newcommand{\bea}{\begin{eqnarray}}
\newcommand{\eea}{\end{eqnarray}}
\newcommand{\bean}{\begin{eqnarray*}}
\newcommand{\ean}{\end{eqnarray*}}
\newcommand{\nn}{\nonumber \\}

\title{Hybrid static potentials and gluelumps on $N_f=3+1$ ensembles}

\author*[a]{Roman~H\"ollwieser}
\author[a]{Francesco Knechtli}
\author[a]{Tomasz Korzec}
\author[b]{Michael Peardon}
\author[a]{Laura Struckmeier}
\author[a]{Juan Andrés Urrea-Niño}
\affiliation[a]{Department of Physics, University of Wuppertal, Gau{\ss}strasse 20, 42119 Germany}
\affiliation[b]{School of Mathematics, Trinity College Dublin, Ireland}

\emailAdd{hoellwieser@uni-wuppertal.de}

\abstract{QCD permits the existence of hybrid mesons that are made up of both quarks and gluons, including exotic states, i.e., quantum numbers prohibited for pure quark-antiquark states, with possible candidates found in experiments. We present static hybrid potentials measured via Laplace trial states together with static-light meson thresholds on $N_f=3+1$ dynamical fermion ensembles with 420 MeV pions. Furthermore, we measure corresponding gluelump masses which refer to the $R\rightarrow0$ limit of the hybrid potentials and are essential input parameters for effective models to describe hybrid mesons.}

\FullConference{The 41st International Symposium on Lattice Field Theory (LATTICE2024)\\
 28 July - 3 August 2024\\
Liverpool, UK\\}

\begin{document}
\maketitle

\section{Introduction}

Static-hybrid quark anti-quark potentials represent a promising frontier in the quest to decipher the strong force's intricacies. These potentials offer a unique perspective on meson behavior, bridging the gap between traditional quark models and lattice QCD simulations~\cite{Dudek:2004epa}. In particular, XYZ states have posed significant challenges to our understanding of hadron spectroscopy. These states do not fit neatly into the quark model, and their properties pose questions about their internal structure and how they relate to QCD~\cite{Swanson:2015wgq,Brambilla:2019esw,Brambilla:2022hhi}. Static-hybrid mesons could provide a theoretical framework to help explain or categorize some XYZ states~\cite{Berwein:2024ztx, Capitani:2018rox}. At small static source separation $r$, the hybrid potentials form multiplets associated with {\it gluelumps}, which are energy levels of QCD in the presence of a color-octet source~\cite{Campbell:1985kp, Jorysz:1987qj}. We show how to construct new hybrid static operators with trial states formed by eigenvector components of the covariant lattice Laplace operator, where the gluonic excitations are realized via covariant derivatives of individual eigenvectors. For the computation of gluelump masses, we follow a classical method~\cite{Foster:1998wu, Marsh:2013xsa, Herr:2023xwg, Herr:2022} refined by using an extended basis of spatial loop shape correlations recently introduced in~\cite{Barca:2024fpc} for glueball calculations.

\section{Methods}\label{sec:meth}

In \cite{Hollwieser:2022doy, Hollwieser:2022pov} we show that we can replace the spatial Wilson line in classical Wilson loops by eigenvector pairs $v_i(\vec x)v_i(\vec y)$ of the three-dimensional gauge-covariant lattice Laplace operator to build so-called {\it Laplace trial state correlators}. We can also write the latter as the spatial correlation of static perambulators $\tau_{ij}(\vec y,t_0,t_1)=v_i^\dagger(\vec y,t_0)U_t(\vec y,t_0,t_1)v_j(\vec y,t_1)$,
and $\bar\tau_{ij}(\vec x,t_0,t_1)$, where $\vec{x}$ and $\vec{y}=\vec{x}+\vec{r}$ denote the location of the static sources. Practically, these static quark line operators are computed first and can be correlated at arbitrary distances, which allows to compute on- and off-axis distances very easily. 

In order to build hybrid Laplace trial states we introduce gluonic excitations via covariant derivatives of the Laplacian eigenvectors
\[\nabla_{\vec k}v(\vec x)=\frac{1}{2}[U_k(\vec x)v(\vec x+\hat k)-U_k^\dagger(\vec x-\hat k)v(\vec x-\hat k)].\]
We extract hybrid static potentials from the correlation functions \begin{eqnarray*}
&&C_{\Sigma_g^+}(R,T)=\sum_{\vec x,t_0,i,j}\big\langle\tr\big[\tau_{ij}(\vec{y},t_0,t_1) \rho^2(\lambda_j) \tau_{ji}(\vec{x},t_1,t_0)  \rho^2(\lambda_i)\big]\big\rangle=\label{eq:CSg}\\
	&&\quad\sum_{\vec x,t_0,i,j}\big\langle\tr\big[U_t(\vec y;t_0,t_1)\rho^2(\lambda_j)v_j(\vec y,t_1)v^\dagger_j(\vec x,t_1)U_t^\dagger(\vec x;t_0,t_1)\rho^2(\lambda_i)v_i(\vec x,t_0)v_i^\dagger(\vec y,t_0)\big]\big\rangle,\nn
 &&C_{\Sigma_{u/g}^\mp}(R,T)=\label{eq:CSi}\\
	&&\quad\sum_{\vec x,t_0,i,j,\vec k||\vec r}\big\langle\tr\big[U_t(\vec y;t_0,t_1)\rho^2(\lambda_j)\{[\nabla_{\vec k}v_j](\vec y,t_1) v^\dagger_j(\vec x,t_1)\pm v_j(\vec y,t_1)[\nabla_{\vec k}v_j]^\dagger(\vec x,t_1)\}\nn
	&&\qquad\qquad\qquad\;
	U_t^\dagger(\vec x;t_0,t_1)\rho^2(\lambda_i)\{[\nabla_{\vec k}v_i](\vec x,t_0)v_i^\dagger(\vec y,t_0)\pm v_i(\vec x,t_0)[\nabla_{\vec k}v_i]^\dagger(\vec y,t_0)\}\big]\big\rangle,\nn
&&C_{\Pi_{u/g}}(R,T)=\label{eq:CPi}\\
	&&\quad\sum_{\vec x,t_0,i,j,\vec k\perp\vec r}\big\langle\tr\big[U_t(\vec y;t_0,t_1)\rho^2(\lambda_j)\{[\nabla_{\vec k}v_j](\vec y,t_1) v^\dagger_j(\vec x,t_1)\pm v_j(\vec y,t_1)[\nabla_{\vec k}v_j]^\dagger(\vec x,t_1)\}\nn
	&&\qquad\qquad\qquad\quad
	U_t^\dagger(\vec x;t_0,t_1)\rho^2(\lambda_i)\{[\nabla_{\vec k}v_i](\vec x,t_0)v_i^\dagger(\vec y,t_0)\pm v_i(\vec x,t_0)[\nabla_{\vec k}v_i]^\dagger(\vec y,t_0)\}\big]\big\rangle,
\end{eqnarray*}
where $R=|\vec r|=|\vec y-\vec x|$ and $T=|t_1-t_0|$.
We improve the overlap of the operator by introducing a set of Gaussian profile functions $\rho_{k}(\lambda_i)=e^{-\lambda_i^2/4\sigma_{k}^2}$ to give different weights to individual eigenmodes $v_i$ according to their eigenvalues $\lambda_i$. We work with seven different Gaussian widths $\sigma_k$ and construct a correlation matrix and solve a generalized eigenvalue problem (GEVP) using the pruned matrix to identify the optimal trial state profiles. 

In the limit of the separation between the $\bm{3}$ and $\bm{3^*}$ sources $r\rightarrow0$, since $\bf 3\,\otimes\,\bf 3^*=\bf 1\,\oplus\,\bf 8$, their effect on the fields of QCD becomes more and more like that of a linear combination of a local color-singlet ({\bf 1}) source and a local color-octet ({\bf 8}) source. While a singlet state decouples from the temporal transporters within an $r=0$ “Wilson loop”, an octet state couples to a temporal Schwinger line in the adjoint representation. In the infinite mass limit, where spin can be neglected, the temporal transporter can be interpreted as the propagator of a static gluon, in analogy to fundamental lines representing a static quark. Hence, only the color-octet component acts as a local source of gluonic fields, restoring the rotational and the discrete symmetries of QCD in the limit $r \to 0$, approaching states with definite $J^{PC}$ quantum numbers and with specific light-quark flavors. In pure $SU(3)$ gauge theory, a state with discrete energy bound to an {\bf 8} source is called a {\it gluelump}~\cite{Alasiri:2024nue, Jorysz:1987qj,Campbell:1985kp}. 
The spectrum of QCD in the presence of a static color-octet source can be calculated using  lattice QCD up to an additive constant~\cite{Foster:1998wu, Marsh:2013xsa, Herr:2023xwg}. We compute gluelump masses from correlation functions of spatial Wilson loops transformed to the adjoint representation connected with an adjoint static line. We perform one HYP2 smearing step for temporal gauge links and three levels of APE smearing steps for spatial links, building a correlation matrix from different levels being treated as a generalized eigenvalue problem (GEVP)~\cite{Blossier:2009kd}. We use $s=1\ldots35$ different 3D loop shapes, denoted as $W_s(\vec x,t)$, starting at a spatial point $\vec x$ at fixed Euclidean time $t$, as introduced in~\cite{Berg:1982kp, Barca:2024fpc} for glueball operators.
Parity is fixed by considering the sum (P=+1) or difference (P=-1) of each loop with its parity twin, {\it i.e.}, the loop under the action of a parity transformation.
Charge conjugation symmetry is fixed by using $W_s^{P,C=\pm}=W_s^P\pm W_s^{P\,\dagger}$.
Elements of the cubic group can act on $W_s^{PC}(\vec x, t)$, resulting in 24 loops which form a basis which generates the regular (or permutation) representation of the cubic group, not considering any degeneracies between loops. Therefore, there exist linear combinations of the 24 loops with projection coefficients transforming according to any of the five irreps of the cubic group $\Lambda\in\{A1,A2,E,T1,T2\}$, were we consider only one copy per irrep. The corresponding coefficients can be calculated via projection methods~\cite{Howard:2018, Bunker:1998} and differ from those for glueballs, since there is no sum over spatial points $\vec x$ and symmetries have to hold locally. We choose projection basis vectors containing no zeros to maximize the use of operators and sum over all loop shapes contributing to a specific irrep, which eliminates the shape index $s$ in the final operators. The projection to the adjoint representation can be avoided by writing the correlator in terms of fundamental operators $W_{\Lambda^{PC}}(\vec x,t)$ and static lines $U_t(\vec x,t_0,t_1)$~\cite{Herr:2023xwg, Herr:2022}
\bea
C_{\Lambda^{PC}}(t)&=&\sum_{\vec x,t_0}\big\{\tr[W_{\Lambda^{PC}}(\vec x,t_0)U_t(\vec x,t_0,t_0+t)W_{\Lambda^{PC}}^\dagger(\vec x,t_0+t)U_t^\dagger(\vec x,t_0,t_0+t)]\\
&&\qquad-\tr[W_{\Lambda^{PC}}(\vec x,t_0)]\tr[W_{\Lambda^{PC}}^\dagger(\vec x,t_0+t)]/3\big\}
\eea
We extract gluelump masses from exponential fits to the correlators containing a self-energy contribution $m_{self}(a)$ from adjoint color sources. The latter differs from the self-energy in the (hybrid) static potentials $V_{self}(a)$, however, the two can be related via~\cite{Foster:1998wu, Bali:2000gf, Bali:2003jq, Bauer:2011ws, Bali:2013pla, Bali:2013qla}
\bea
m_{self}(a)=\frac{C_A}{C_F}\frac{V_{self}(a)}{2}=\frac{N^2}{N^2-1}V_{self}(a)>V_{self}(a),\label{eq:glse}
\eea
with $C_A=N$ and $C_F=(N^2-1)/2N$ the quadratic Casimir invariants of the adjoint and fundamental representations of $SU(N)$. Since for $SU(3)$ the relative factor is close enough to one, we include the gluelump masses in the potential plots without additional shifts.

\section{Ensembles}\label{sec:ens}

We work with one quenched ensemble (qE), an $N_f=2$ ensemble (Em1) with two dynamical quarks of half the charm quark mass ($m_\pi=2.2$ GeV), both of lattice size $48\times24^3$ and $N_f=3+1$ ensembles with 788 MeV (A1h) and 406 MeV (A1) pions on $96\times32^3$ lattice volumes, as well as the latter pion mass also on a larger volume (A2, $128\times48^3,\,m_\pi=409$ MeV)~\cite{Hollwieser:2020qri}. The $48\times 24^3$ lattices have periodic boundary conditions while the $N_f=3+1$ ensembles have open boundaries in temporal direction. They were all produced with the openQCD package \cite{Luscher:2012av} using the plaquette (qE,Em1) and Lüscher-Weisz gauge action (A1h,A1,A2). For the latter ($N_f=3+1$) ensembles we use Wilson quarks with a non-perturbatively determined clover coefficient in a massive O(a) improvement scheme~\cite{Fritzsch:2018kjg}. The bare gauge couplings $g_0^2=6/\beta$ and the hopping parameters, as well as the Sommer scale $r_0/a$~\cite{Sommer:1993ce}, the flow scale $t_0/a^2$~\cite{Luscher:2010iy} and corresponding lattice spacing (defined via the mass difference of $h_c$ and $\eta_c$ \cite{Hollwieser:2020qri}) are given in \tab{tab:latt}. All measurements were performed by our C+MPI based library that facilitates massively parallel QCD calculations. A total of $N_{vec}$ eigenvectors of the 3D covariant Laplacian were calculated on each time-slice of the lattices. 
A total of 20 3D APE smearing \cite{Albanese1987} steps with $\alpha_{APE} = 0.5$ were applied on each gauge field before the eigenvector calculation so as to smooth the link variables that enter the Laplacian operator. When forming the correlations of the Laplace trial states, we apply one HYP2 smearing step with parameters $\alpha_1=1,\,\alpha_2=1$ and $\alpha_3=0.5$ to the temporal links~\cite{Hasenfratz:2001hp, DellaMorte:2003mw, DellaMorte:2005nwx, Grimbach:2008uy, Donnellan:2010mx}, which corresponds to a particular choice of the static action. 
The error analysis in this work was done using the $\Gamma$ method \cite{Wolff:2003sm,Schaefer:2010hu} in a recent python implementation (pyerrors)~\cite{Joswig:2022qfe} with automatic differentiation~\cite{Ramos:2020scv}.

\begin{table}[h!]
  \centering
\begin{tabular}{|c||c|c|c|c|c|}
\toprule
ensemble & qE & Em1 & A1h & A1 & A2 \\
\midrule
volume & $48\times24^3$ & $48\times24^3$ & $96\times32^3$ & $96\times32^3$ & $128\times48^3$ \\
$\beta$ & 5.85 & 5.3 & 3.24 & 3.24 & 3.24 \\
$\kappa_l$ & - & - & 0.13392 & 0.134407 & 0.134396\\
$\kappa_c$ & - & 0.13270 & 0.12834 & 0.12784 & 0.12798 \\
$m_\pi$ & - & 2.2(1) GeV & 788(6) MeV & 406(3) MeV & 409(1) MeV \\
$r_0/a$  & 4.2612(42) & 4.2866(24) & 7.278(27) & 9.023(63) & 8.998(54) \\
$t_0/a^2$ & - & 1.8477(3) & 5.078(14) & 7.438(32) & 7.434(20) \\
$a$ [fm] & 0.0662(12) & 0.0658(10) & 0.0690(11) & 0.05359(15) & 0.05355(13) \\
\midrule
$N_{cfg}$ & 6480 & 7098 & 4000 & 4000 & 2000 \\
$N_{vec}^l$ & - & - & 200 & 100 & 100 \\
$N_{per}^l$ & - & - & 2000 & 4000 & 2000 \\
$N_{vec}^c$ & - & 200 & 200 & 200 & - \\
$N_{per}^c$ & - & 1000 & 2000 & 4000 & - \\
\bottomrule
\end{tabular}
\caption{\label{tab:latt}Lattice, simulation and distillation (numbers of eigenvectors and perambulators for light and charm quarks) parameters of the ensembles qE (quenched), Em1, A1 heavy and A1 resp. A2 (light).} 
\end{table}

\section{Results}\label{sec:res}

\begin{figure}
\centering
\includegraphics[width=.9\linewidth]{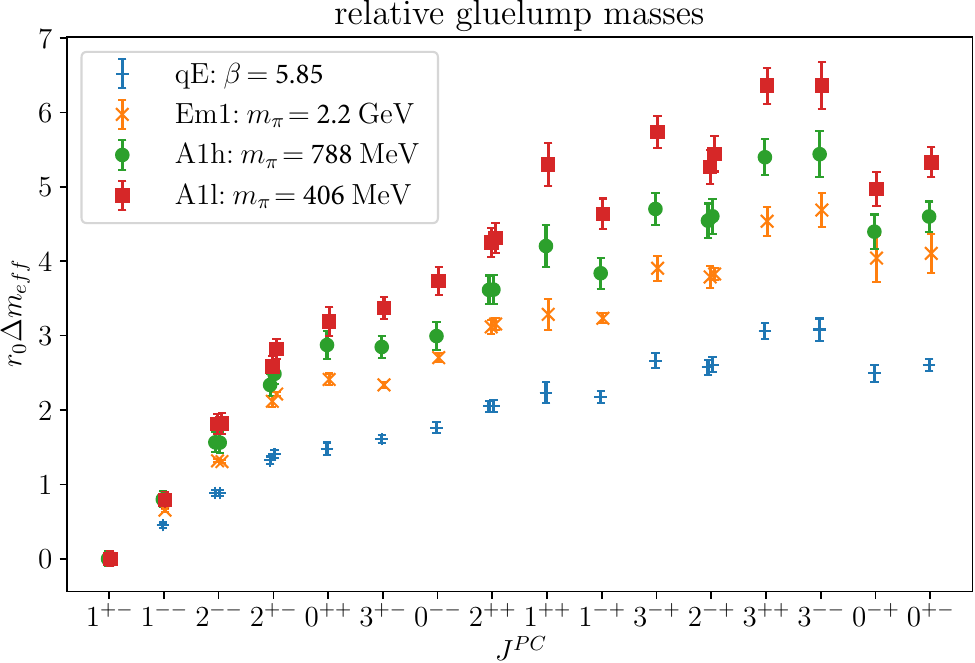}
\caption{Relative gluelump masses $r_0\Delta m_{eff}=r_0/a\,(\,am_{eff,\Lambda}-am_{eff,T1^{+-}}\,)$ on ensembles qE (quenched), Em1 ($m_\pi=2.2$GeV), A1 heavy ($m_\pi=788$ MeV) and A1 light ($m_\pi=406$ MeV).}
  \label{fig:gls2}
\end{figure}

We present gluelump masses measured on the full statistics on all ensembles. Since the lattice energies contain the unphysical self-energy \Eq{eq:glse}, only mass differences have a continuum limit and we consider the mass differences with respect to the $T^{+-}$ state, which happens to be the lowest state. We follow the ordering of low lying gluelumps as been established~\cite{Foster:1998wu} and plot the masses against $J^{PC}$ assuming the lowest spin contained in the $O_h$ representation in \fig{fig:gls2} for various ensembles. The results on the two similar ensembles A1 and A2 only differing by their volumes are in perfect agreement with each other, hence we do not include the masses of A2 in this plot. It is also reassuring to see that $E^{PC}$ and $T_2^{PC}$ are consistent with each other for each PC, since they should couple to the same physical state ($J=2$) in the continuum limit, hence we plot them with a slight offset next to each other in the corresponding $2^{PC}$ channel. We find the individual energies consistently increasing with decreasing pion mass.

\begin{figure}
\centering
\includegraphics[width=.9\linewidth]{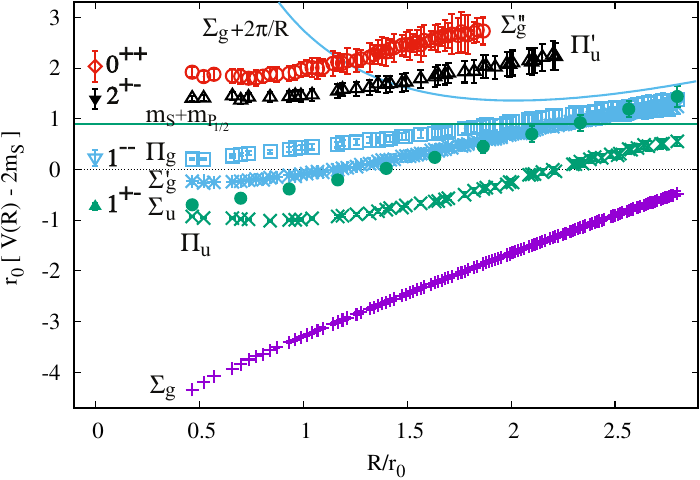}
\includegraphics[width=.9\linewidth]{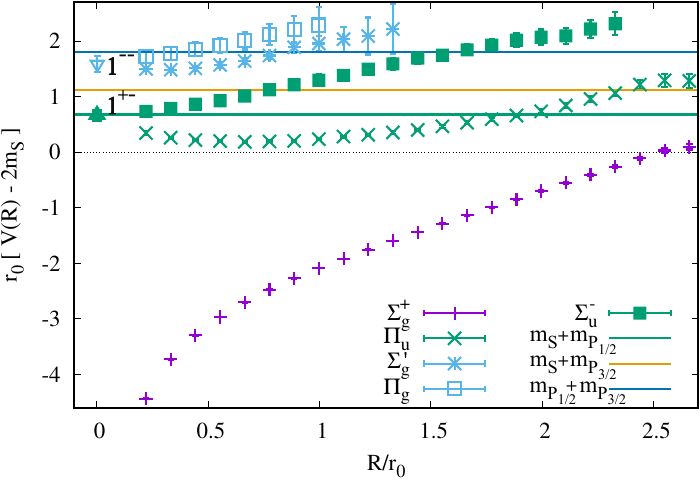}
\caption{Hybrid static potentials and corresponding gluelumps on Em1 (top) and A2 (bottom) relative to twice the static-light S-wave mass $m_S$ and the $P_{1/2}$-wave mass $m_{P_{1/2}}$. For A2 we also show the $P_{3/2}$-wave mass.}\label{fig:states}
\end{figure}

We show the hybrid static ground state potentials of $\Sigma_{g/u}$ and $\Pi_{g/u}$ and some excited states in \fig{fig:states} for Em1 and A2 ensembles. We plot the potentials relative to twice the static-light S-wave energy $m_S$ and also mark the static light $P_{1/2}$ and $P_{3/2}$-wave energies $m_{P_{1/2}}$ and $m_{P_{3/2}}$. We see that for Em1 (and A1) expected string breaking distances of $\Sigma_g$ and $\Pi_u$ are just above half the spatial lattice extent, on A2 we could actually study hybrid string breaking. For on-axis separations the potential of $\Pi_{g/u}$ in the continuum representation can be obtained from the $E_1^\mp$ representation of $D_{4h}$. For off-axis separations (Em1 only) we technically do not have $D_{4h}$, but we can consider off-axis separations in a 2D plane only rather than the 3d volume, to be left with one orthogonal direction for the covariant derivatives, see also~\cite{Bali:2000vr}. For $\Sigma_u$ with derivatives along the separation axis we compute on-axis distances only. We include the lowest gluelump masses corresponding to the $R\rightarrow0$ limit of the lowest hybrid potentials, also subtracting twice the static-light S-wave mass, which does not completely remove the self-energy contribution. Nevertheless, the limits look reasonable. 

\section{Conclusions \& Outlook}\label{sec:co}

We have computed gluelump spectra and hybrid static potentials $V_{\Lambda_\eta^\epsilon}(r)$ for $\Lambda_\eta^\epsilon = \Sigma_{g/u}$ and $\Pi_{g/u}$ states in SU(3) lattice gauge theory. 
For the computation of gluelump masses we use an extended basis of spatial loop shape correlations. Comparison of $E$ and $T_2$ representations, both of which couple to $J=2$ in the continuum limit, provides a cross-check on cutoff effects. For the hybrid static potentials we use alternative operators for a static quark-anti-quark pairs based on Laplacian eigenmodes, replacing traditional Wilson loops. Instead of "gluonic handles" (excitations) of the spatial Wilson lines we use symmetric, covariant derivatives of the Laplacian eigenvectors to form improved Laplace trial states by applying optimal profiles to give different weights to individual eigenvectors, derived from a generalized eigenvalue problem. A high resolution of the hybrid static potentials can be achieved as off-axis distances can easily be computed in the new approach. We present a hybrid static spectrum including excited states and show their optimal profiles which allow us to extract effective energies from reasonable effective mass plateaus. In the spectrum we also mark the string breaking thresholds from static-light S- and P-waves, on ensemble A2 we could actually study the string breaking of $\Sigma_g$ and $\Pi_u$. The computation of (hybrid) static-light correlators using "perambulators" $v(t_1)D^{-1}v(t_2)$ from the distillation framework was presented in~\cite{Struckmeier:2025ebs}. This allows us to put together the building blocks for string breaking in QCD, computing the mixing matrix including static and light quark propagators. The new methods can also be applied to tetra- and multi-quark potentials, which is the plan for future investigations.

\section*{Acknowledgements} The authors gratefully acknowledge the Gauss Centre for Supercomputing e.V. (www.gauss-centre.eu) for funding this project by providing computing time on the GCS Supercomputer SuperMUC-NG at Leibniz Supercomputing Centre (www.lrz.de). M.P. was supported by the European Union’s Horizon 2020 research and innovation programme under grant agreement 824093 (STRONG-2020). The work is supported by the German Research Foundation (DFG) research unit FOR5269. 
The project 
received funding from the programme " Netzwerke 2021", an initiative of the Ministry of Culture and Science of the State of Northrhine Westphalia, in the NRW-FAIR network, funding code NW21-024-A (R.H.). We thank Abhishek Mohapatra, Nora Brambilla and Stefano Capitani for valuable discussions.

\bibliographystyle{JHEP}
\bibliography{paper}

\end{document}